\documentclass[]{raa}
\usepackage{bbm}            % referee version: for submission
\usepackage{graphicx,times}
\usepackage{natbib}

\providecommand{\bm}[1]{\mbox{\boldmath$#1$\unboldmath}}

\begin{document}

   \title{Quasi-period outflows observed by the X-Ray
   Telescope onboard \textit{Hinode} in the boundary of an active region}

 \volnopage{ {\bf 2010} Vol.\ {\bf 0} No. {\bf XX}, 000--000}
   \setcounter{page}{1}

   \author{L.-J. Guo
      \inst{1}
   \and H. Tian
      \inst{1}
   \and J.-S. He
      \inst{2}
   }

   \institute{School of Earth and Space Sciences, Peking University,
             Beijing 100871, China; {\it guolijai32@pku.edu.cn}\\
        \and
             Max-Planck-Institut f\"ur Sonnensystemforschung, 37191 Katlenburg-Lindau, Germany
}

\abstract{Persistent outflows have recently been detected at
boundaries of some active regions. Although these outflows are
suggested to be possible sources of the slow solar wind, the nature
of these outflows is poorly understood. Through an analysis of an
image sequence obtained by the X-Ray Telescope onboard the
\textit{Hinode} spacecraft, we found that quasi-period outflows are
present in the boundary of an active region. The flows are observed
to occur intermittently, often with a period of 5-10 minutes. The
projected flow speed can reach more than 200~km/s, while its
distribution peaks around 50~km/s. This sporadic high-speed outflow
may play an important role in the mass loading process of the slow
solar wind. Our results may imply that the outflow of the slow solar
wind in the boundary of the active region is intermittent and
quasi-periodic in nature. \keywords{ Solar wind, UV radiation,
Active region } }

   \authorrunning{Guo et al. }            %author_head in even pages
   \titlerunning{Quasi-period outflows observed by XRT}  % title_head in odd pages
   \maketitle

%
%________________________________________________ sections below
%
\section{Introduction}           %% first-level sections will be auto-capitalized
\label{sect:intro}

%solar wind
Solar wind origin is among the most important unresolved problems in space and solar physics,
although a lot of investigations have been performed in this field. It is commonly believed that
coronal holes are the source regions of the fast solar wind
\citep[e.g.,][]{Krieger1973,Hassler1999,Tu2005,Tian2010}. While there exist several candidates for
the source of the slow solar wind, such as helmet streamers and boundaries of polar coronal holes
\citep[e.g.,][]{Wang1990,Chen2004,Antonucci2006,Kohl2006}, local open-field regions in the quiet
Sun \citep[e.g.,][]{He2007,Tian2008a,Tian2009,He2009}, and boundaries of some active regions (ARs)
\citep[e.g.,][]{Kojima1999,Sakao2007,Marsch2008,Harra2008}.

%mass flow
Outflows at edges of some ARs were previously observed in both
imaging and spectroscopic observations. \citet{Winebarger2001}
reported outflows from an AR observed on 1 December 1998. The
velocities of the outflows in coronal loops were estimated as
between 5 and 20 km/s. The authors postulated that these mass flows
are driven by small-scale magnetic reconnection events occurring at
the foot points of coronal loops. Using observations made by the
X-Ray Telescope (XRT) \citep{Golub2007} onboard the \textit{Hinode}
spacecraft, \citet{Sakao2007} identified outflows with a speed of
$\sim$140 km/s from the edge of an AR. With the help of coronal
magnetic field extrapolation from the photospheric magnetogram, they
found that these outflows are probably associated with open field
lines and thus suggested that these outflows might correspond to the
slow solar wind. This conclusion was supported by \citet{Harra2008}
and \citet{Marsch2008}, who found that prominent blue shifts of the
coronal emission line Fe\,{\sc xii}~195~{\AA} are associated with
open field lines in AR boundaries. \citet{DelZanna2008} further
found that the outflows in AR boundaries are temperature dependent
and the speed of the outflows increases steadily with increasing
temperature.

%wave
Quasi-periodic intensity fluctuations have been identified in different parts of the Sun. These
intensity oscillations are usually interpreted as disturbances caused by slow magnetoacoustic waves
with different periods. In ARs, quasi-periodic oscillations and small-scale propagating
brightenings are often found to be associated with coronal loops
\citep[e.g.,][]{DeMoortel2000,DeMoortel2002,Robbrecht2001}. The low-frequency oscillations could be
the result of the leakage of the slow magnetoacoustic waves to the corona along inclined magnetic
filed lines from the lower atmosphere \citep[e.g.,][]{DePontieu2005}, or recurrent magnetic
reconnections between large-scale loops and small loops in ARs \citep[e.g.,][]{Baker2009}.

%describe
In this paper, we apply a wavelet analysis to the outflow events in
the boundary of an AR by using high-cadence XRT observations. The
outflows, which seem to be associated with open magnetic field
lines, are found to be sporadic and reveal a clear characteristic of
periodic occurrence.

\section{Observation and Data analysis}
\label{sect:Obs}

%----------Figure.1
\begin{figure}
\centering
\includegraphics[width=140mm]{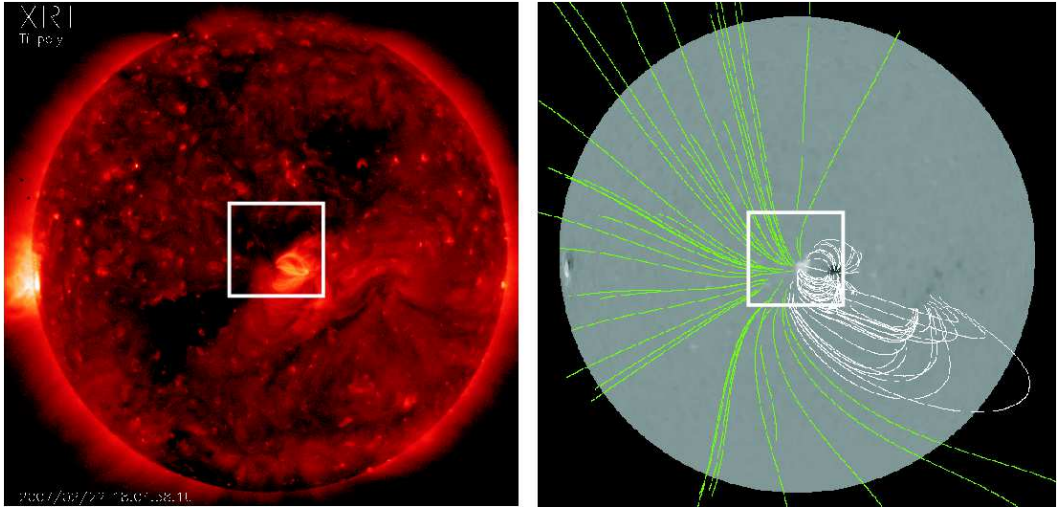}
\caption{ Left: the full Sun image taken with XRT at 18:04:38 on
February 22, 2007. Right: magnetic field lines in AR NOAA 10942 at
18:04:00 on February 22, 2007, obtained from an extroplation from
the full-disk photospheric magnetogram of GONG (Global Oscillation
Network Group) by using the PFSS package of SSW. Green and white
lines represent open and closed field lines, respectively. The white
rectangle in each panel marks the field of view shown in
Figure~\ref{fig.2}.}
 \label{fig.1}
\end{figure}

%----------Figure.2
\begin{figure}
\includegraphics[width=73mm]{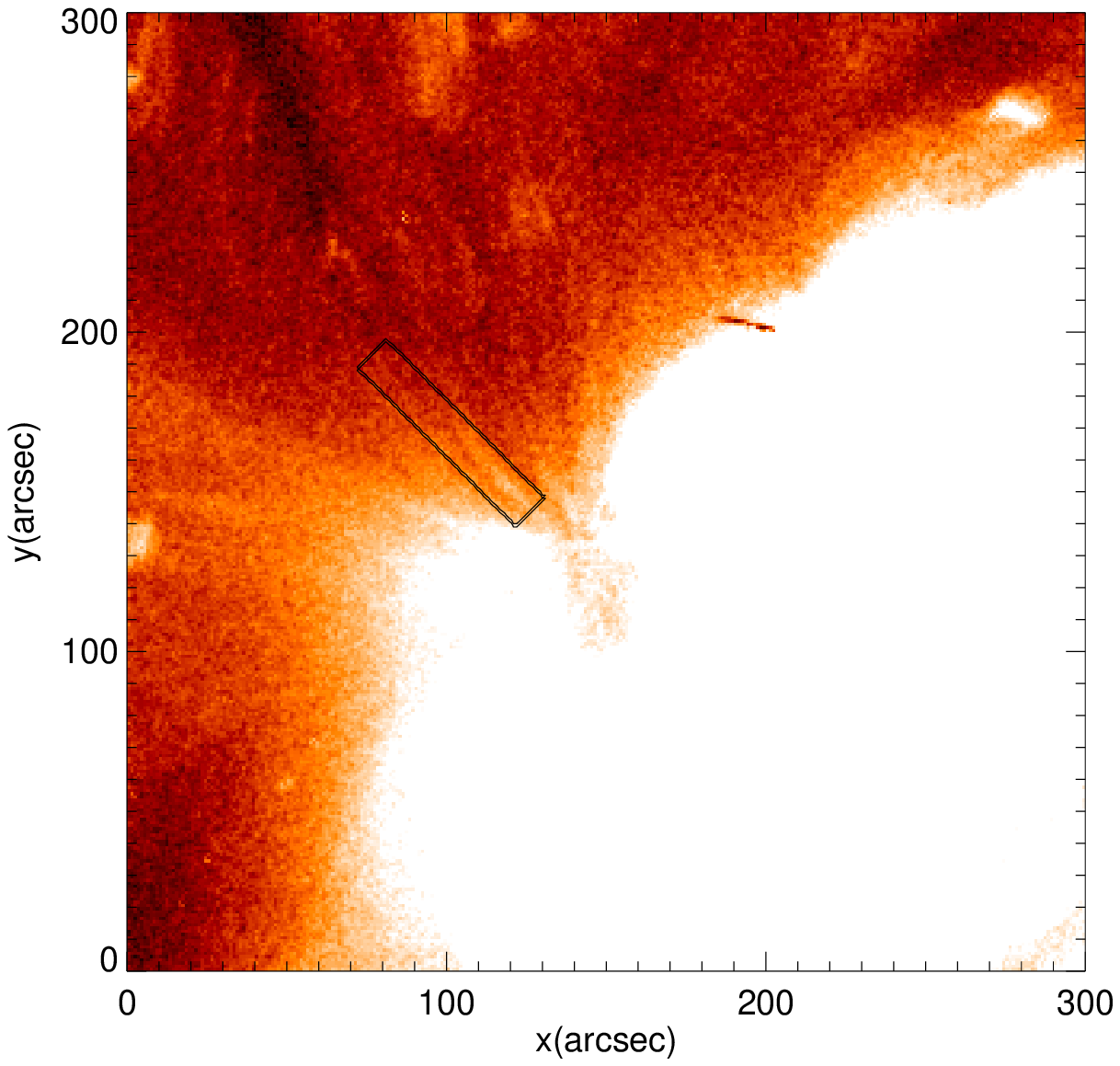}
\includegraphics[width=73mm]{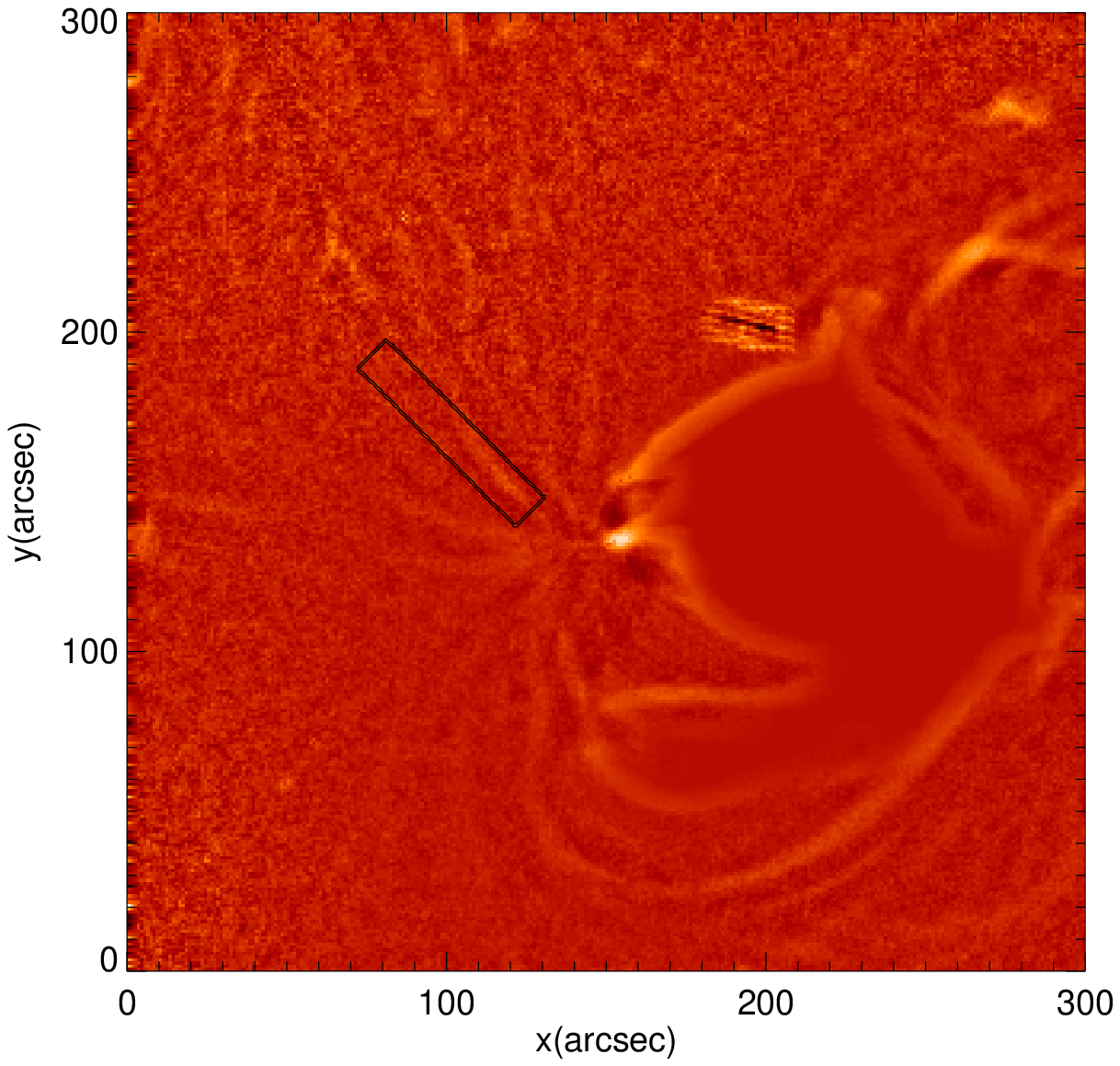}
\caption{An image of the AR NOAA 10942 obtained with the XRT Ti-poly
filter at 11:59:33 UT on February 22, 2007. Left: the level-1 image.
Right: the contrast-enhanced image. The black rectangle in each
image marks the strand-like structure for which a wavelet analysis
is performed. }
   \label{fig.2}
\end{figure}

The data set used in this article was obtained by the XRT instrument
onboard \textit{Hinode} in AR NOAA 10942. An XRT Ti-poly image
sequence of the AR NOAA 10942 was obtained from 11:33 to 21:27 on
February 22, 2007, including 335 images with a cadence of 90 s. The
spatial resolution was 2 arcsecs and the exposure time was 16~s. The
IDL routine \textit{xrt\_prep.pro} was applied to correct and
calibrate the level-0 XRT data to level-1 data. The full Sun image
taken with XRT at 18:04 on February 22, 2007 is shown in
Figure~\ref{fig.1}. Right panel of Figure~\ref{fig.1} shows the
full-disk photospheric magnetogram of GONG (Global Oscillation
Network Group) with the projection of the extrapolated coronal
magnetic field. The magnetic field lines in the target AR is
obtained by using the potential field source surface (PFSS) model
\citep[e.g.,][]{Schrijver2001}. It is very clear that the dark
coronal-emission region on the eastern side of the AR is associated
with open field lines, and thus is possible to be a source region of
the solar wind outflow. A level-1 image at 11:59:33 UT is shown in
the left panel of Figure~\ref{fig.2}. The intensity contrast in this
image was sharpened and the result is presented in the right panel
of Figure~\ref{fig.2}. From Figure~\ref{fig.2} we can identify some
plume-like or strand-like enhanced emission structures inside the
reduced-emission region.

%--------------Figure.3
   \begin{figure}
   \centering
   \includegraphics[width=140mm]{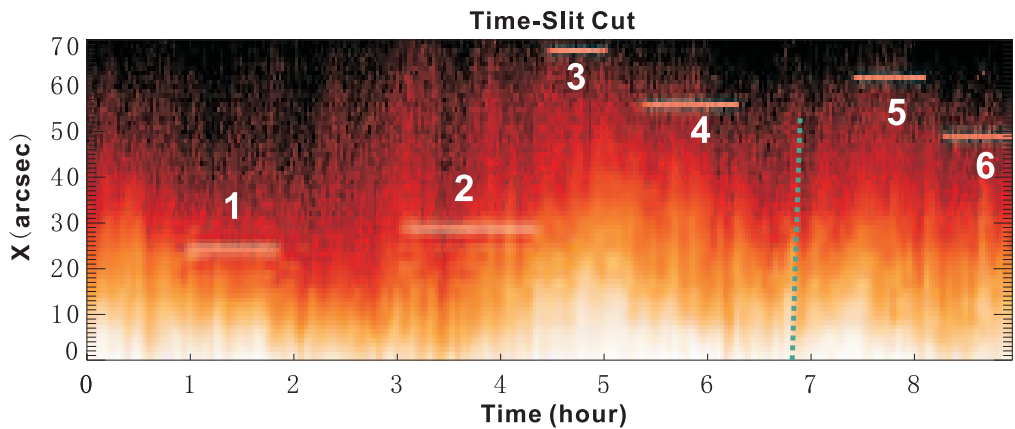}
   \caption{ The temporal evolution of the intensity along the strand.
   The vertical axis represents positions
   along the chosen strand as illustrated in Figure~\ref{fig.2}. The six white horizontal lines mark
   six intervals of the intensity fluctuation for which wavelet analyses have been performed. The
   green dot line is placed along the enhanced intensity caused by one outflow event.}
   \label{fig.3}
   \end{figure}

We selected one strand-like structure, which is marked by the black
rectangle in Figure~\ref{fig.2}, for in-depth analysis. We did not
include the root of strand and its far end, in order to avoid
interference with other structures near the two ends. First the time
sequence of the X-Ray emission in this rectangular region was
extracted. We defined the long side of the rectangle to be the
X-direction, which is approximately parallel to the orientation of
the strand. The short side of the rectangle was defined to be the
Y-direction and it is perpendicular to the strand orientation. At
each time, we then averaged the intensities over the Y-range at each
X position and obtained the average intensity along the strand. We
derived the average intensity along the strand and then constructed
a time-height diagram revealing the temporal evolution of the
intensity along the strand. This time-height diagram is presented in
Figure~\ref{fig.3}.

From Figure~\ref{fig.3} we can see that enhanced intensity features
(outflows seen in the movie) appear frequently. Some features
attenuate quickly when reaching far away from the root of the stand.
But some others can still be clearly identified at the far end of
the strand. The occurrence of the events of intensity enhancement
seems to be sporadic. In order to study the possible recurrent
nature of the outflows, we performed wavelet analyses for the time
series at some positions along the strand. The wavelet analysis of a
time series can reveal both the periods of fluctuations and how the
oscillation with a specific period varies in time
\citep{TorrenceCompo1998}, and thus it's a very useful tool to probe
the sporadic behavior of the outflow here. We randomly chose six
time series, with each observed at different position of the strand
within different time range. The six intervals of the intensity
fluctuation are marked by the six white horizontal lines in
Figure~\ref{fig.3}.

%---------Figure.4
\begin{figure}

\centering
\includegraphics[width=146mm]{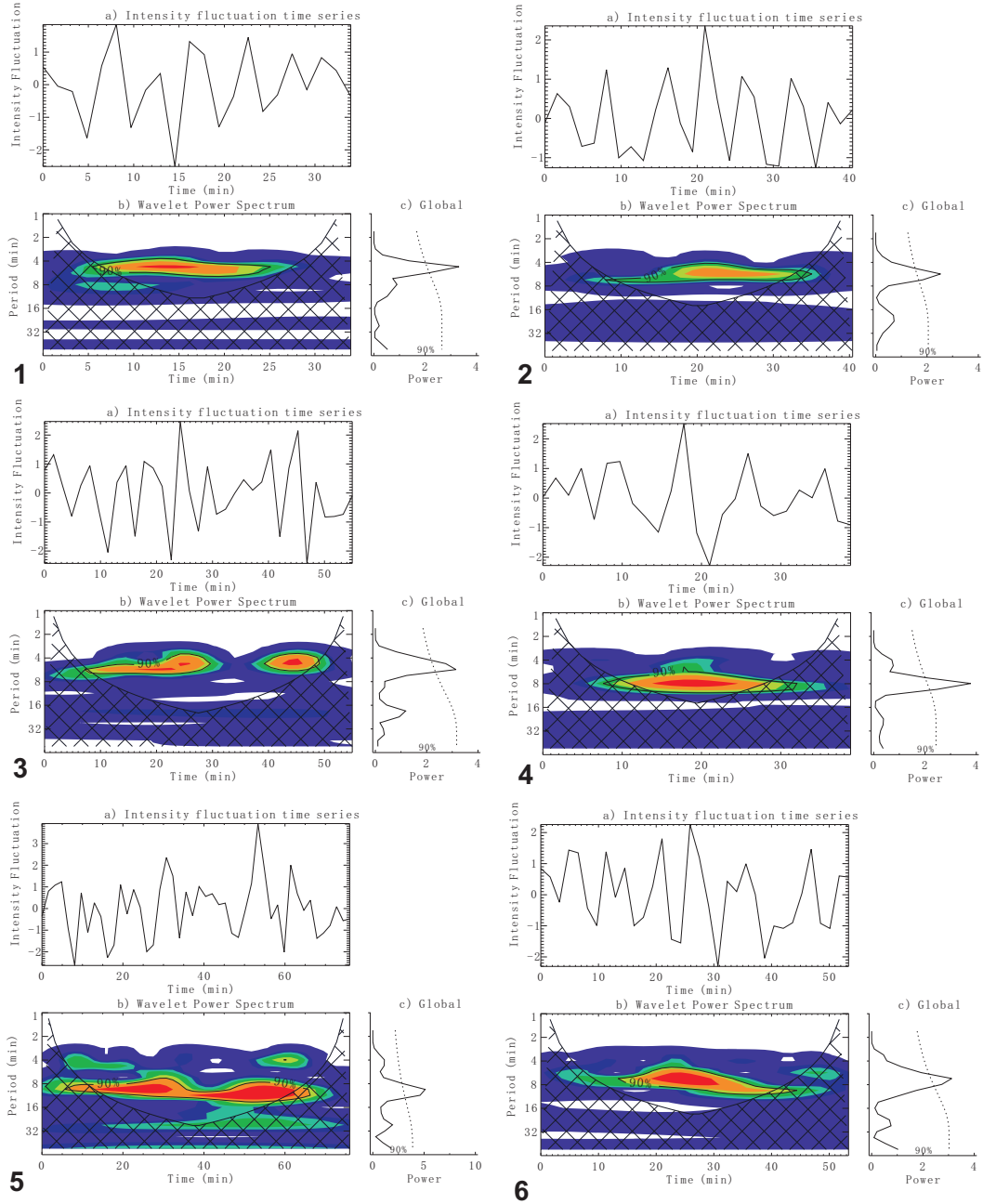}
\caption{~Wavelet power spectra of the six time series as illustrated in Figure~\ref{fig.2}. (a)
The intensity fluctuation observed by the XRT instrument. (b) Time/period variation of the wavelet
power spectrum. Cross-hatched regions indicate the $^{\prime\prime}$cone of
influence$^{\prime\prime}$. The red and blue parts represent the largest and smallest power,
respectively. The contours correspond to the 90\% confidence level. (c) Global wavelet. The dotted
lines correspond to the 90\% confidence level.}
   \label{fig.4}
\end{figure}

Here we also chose the Morlet wavelet function for our analysis. The
Morlet wavelet function is defined as a sine wave modulated by a
Gaussian window. This wavelet function has been widely used for
analyses of oscillations found in different solar features
\citep[e.g.,][]{Banerjee2000,Tian2008b,Chen2008}. Each time series
of the intensity fluctuation, its wavelet power spectrum, as well as
the global wavelet spectrum are shown in Figure~\ref{fig.4}. The
colors in the wavelet power spectrum represent the relative
intensity of the power spectrum, with the red and blue being the
largest and smallest, respectively. A significance test was also
performed to check if the periodic signatures revealed in the
wavelet power spectrum are real or not. Here we chose confidence
level of 90\%. As mentioned by \cite{TorrenceCompo1998}, wavelet
transform has the edge effect at both ends of the time series. The
influence range of the edge effect is outlined by the cross-hatched
region (cone of influence) in each wavelet power spectrum. The
periodic signatures in this cross-hatched region are difficult to be
determined as real or not. The global wavelet spectrum can be
regarded as an average of the wavelet power spectrum over the time
domain. For more information on the wavelet analysis, we refer to a
complete description by \cite{TorrenceCompo1998}.

%----------Figure.5
\begin{figure}
\centering
\includegraphics[width=100mm,height=90mm]{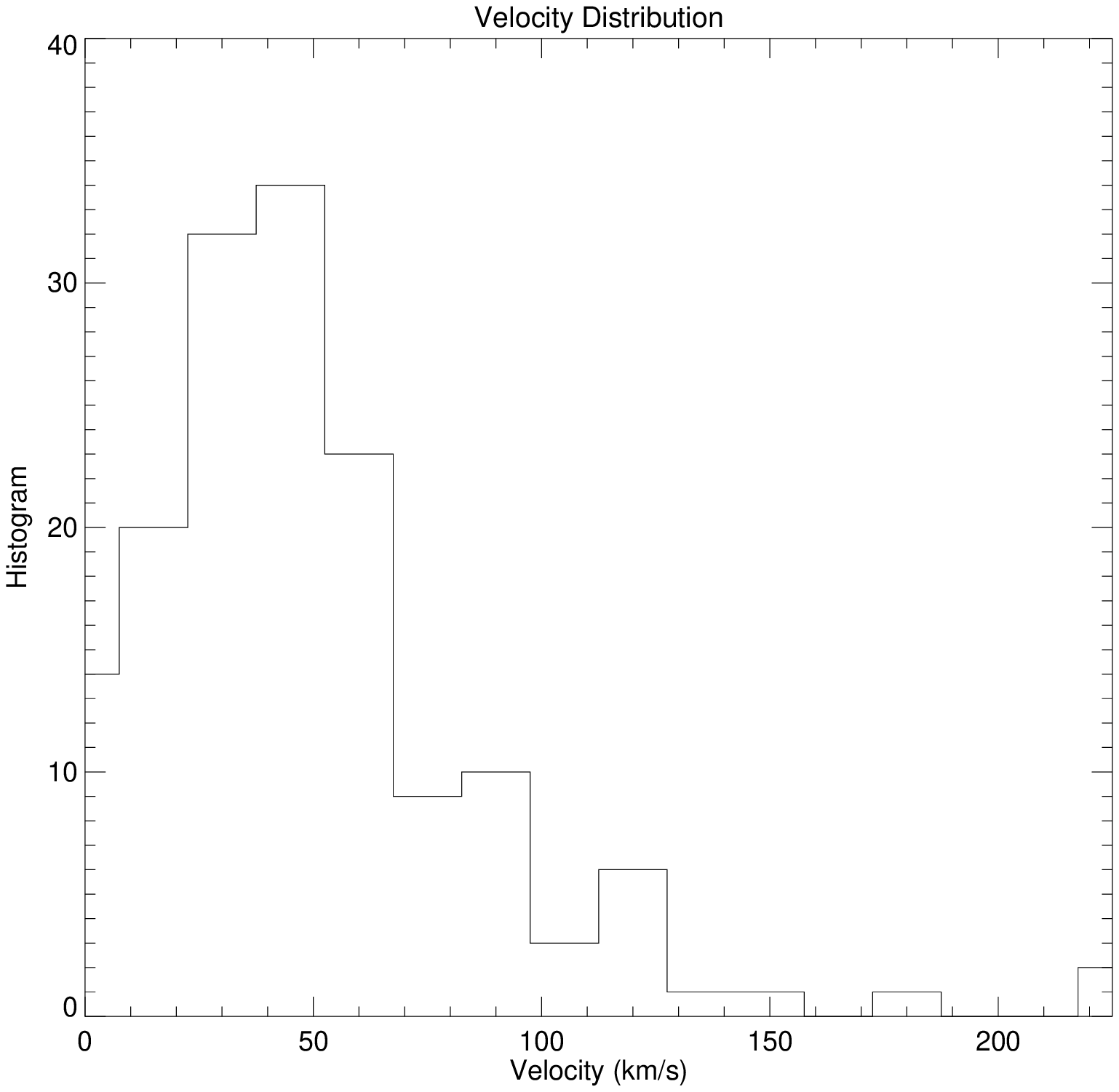}
\caption{This figure shows the distribution of the outflow
velocities in strand-like structures. The X axis is the velocity
range of all events, which is divided into 15 intervals. The Y axis
is the number of outflow events which fall into these intervals.}
   \label{fig.5}
\end{figure}

We also estimated the velocities of the outflows. From the image
sequence we used here, we identified 156 events occurring
respectively in 8 strand-like structures. Different strands extend
along different directions, perhaps depicting the different
stretching directions of local magnetic field lines at the edge of
the active region. The projected velocity of each outflow event was
estimated from the slope of the corresponding intensity enhancement.
The green dot line in Figure~\ref{fig.3} marked the trend of the
intensity enhancement caused by one outflow event, and is used as an
example to calculate the outflow velocity. This method has been
frequently adopted to derive the projected velocities of directed
flows or propagating disturbances
\citep[e.g.,][]{Sakao2007,Tian2008b,McIntosh2010}. We use this
method to calculate the projected velocities of the 156 outflow
events in strand-like structures and present their distribution in
Figure~\ref{fig.5}.

%%%%%%%%%%%%%%%%%%%%%%%%%%%%%%%%%%%%%%%%%%%%%%%%%%%%%%%%%%%%%%
%%     Examples for figures using graphicx for LaTeX 2e
%%               -- our recommended way for embodying graphics
%%%%%%%%%%%%%%%%%%%%%%%%%%%%%%%%%%%%%%%%%%%%%%%%%%%%%%%%%%%%%%

\section{Results and Discussion}
\label{sect:discussion}

Outward propagating disturbances or intensity enhancements in ARs were studied by several authors.
However, most of these studies found that these quasi-period oscillations are associated with
coronal loops in ARs \citep[e.g.,][]{DeMoortel2000,DeMoortel2002,Robbrecht2001}. The low-frequency
oscillations found in the corona are suggested to result from the upward leakage of the slow
magnetoacoustic waves along inclined magnetic filed lines \citep[e.g.,][]{DePontieu2005}, or
recurrent magnetic reconnections between large-scale loops and small cool loops
\citep[e.g.,][]{Baker2009}.

From Figure~\ref{fig.1}, we find that the boundary of the AR studied
here is more likely to be an open magnetic field region rather than
a closed field region. Recent studies using magnetic field
extrapolations also demonstrate clearly that such kind of
weak-emission regions are associated with open field lines extending
outward to the interplanetary space
\citep{Sakao2007,Marsch2008,Harra2008,He2010}. So these outflows are
probably corresponding to the outflow of the slow solar wind from AR
edges. The presence of the significant blue shift of the coronal
line Fe\,{\sc xii}~195~{\AA} in the same regions provides further
support for this conclusion
\citep{Marsch2008,Harra2008,DelZanna2008}.

The strand-like structures are probably the footpoints of the open field lines, and we mainly
observed the outward propagating disturbances along open field lines in the boundary of the AR.
Although we can not exclude the possibility that these outward propagating disturbances could be
the manifestation of upward propagating slow magnetoacoustic waves, the obvious link between the
solar wind outflow and the edge of the AR seems to suggest that the propagating intensity
enhancement might be the exhibition of the outflow of the slow solar wind along open field lines
\citep{He2010}.

Figure~\ref{fig.2} reveals that the outflow is not really continuous, but sporadic in time. And it
seems in the image that the dynamic events occur from time to time and follow a certain frequency.
The wavelet analyses in Figure~\ref{fig.3} show clear periods of the intensity fluctuation ranging
from 5 to 10 minutes. The periods seem to be different within different time intervals.

These results may have important implications of the nature of the
solar wind origin. Almost all of the solar wind model assume that
the solar wind flows continuously from the source regions. However,
our results reveal clearly that the possible initial outflow of the
slow solar wind is intermittent and quasi-period, and thus put more
observational constraints on future solar wind models.

The intermittent and periodic nature of the outflow may result from
intermittent dynamic processes in the lower solar atmosphere. Our
recent study of \citet{He2010} suggested that chromospheric dynamics
might play an important role in the formation of the solar wind in
AR boundaries. A 5-minute oscillation was found in the underlying
chromospheric emission in that paper. Our present study may further
suggest a link between the chromospheric quasi-periodic disturbance
and the quasi-periodic coronal outflow. Magnetic reconnections
between the open magnetic field lines and small-scale cool loops in
the boundary of the AR are likely to be the driver of the outflows.
Reconnections may occur when the frozen-in condition is locally
destroyed, while the reconnection process will switch off when this
condition is restored. As a result, the recurrent reconnection
processes will drive the outflows quasi-periodically and
intermittently. The photospheric quasi-periodic oscillation might
also play a role in driving the outflows. As the oscillation
propagate into the upper solar atmosphere along the highly inclined
magnetic field lines in the AR boundary, acoustic shocks might form
to elevate the local materials \citep{DePontieu2004}, thus could
also play an important role in heating and driving the materials
into the corona and form the solar wind.

The distribution of the outflow velocities presented in
Figure~\ref{fig.5} peaks around 50~km/s. We have to mention that
these velocities are only the projected components of the real
velocities. Many of the magnetic field line in the boundary of the
AR are in fact highly inclined and thus making a angle to the plane
of the spy. So the real outflow speed could be even larger.
Recently, \citet{McIntosh2010} identified high-speed propagating
intensity enhancement in polar plumes with a mean velocity of
135~km/s. The authors concluded that these outflows may originate in
the upper chromosphere and transition region, and could play an
important role in the mass loading process of the fast solar wind.
The ubiquitous high-speed outflows along the strand-like structures
in AR boundaries might be the phenomena similar to those identified
by \citet{McIntosh2010}, and could thus be essential in the mass
loading process of the slow solar wind.

We are also aware that the magnetic field configuration shown in
Figure~\ref{fig.1} might not be accurate due to the defects of the
potential field model, although the PFSS model is widely used and
accepted. If the dark region on the eastern side of the AR is
associated with closed field lines, the intermittent outflows we
observed could be considered either mass supply to the very large
corona loops \citep{Tian2008a,Tian2009}, or being related to the
mass heating and injection events originating from the chromosphere
\citep{McIntosh2009}.

\section{Summary}
\label{sect:conclusion}

We have performed a wavelet analysis for the coronal outflows in the
boundary of an active region. The flows are observed to be
intermittent and often exhibit periods ranging from 5 to 10 minutes.
Statistic study shows that the distribution of the projected
component of the flow speed peaks around 50~km/s. Based on
extrapolated local open field lines and previous findings that the
dark-emission regions in boundaries of some ARs exhibit significant
blue shift of the coronal line Fe\,{\sc xii}~195~{\AA}, we suggest
that the high-speed outflows may play an essential role in the
process of the mass loading of the slow solar wind and that the
sporadic quasi-periodic outflows might result from intermittent
small-scale magnetic reconnections in the chromosphere and
transition region.

\normalem
\begin{acknowledgements}
XRT is an instrument onboard {\it Hinode}, a Japanese mission
developed and launched by ISAS/JAXA, with NAOJ as domestic partner
and NASA and STFC (UK) as international partners. It is operated by
these agencies in co-operation with ESA and NSC (Norway). The
authors are supported by the National Natural Science Foundation of
China under contracts 40874090 and 40931055.
\end{acknowledgements}

%\appendix                  %%appendicial material is supported

%\section{This shows the use of appendix}
%A postscript file is actually an ASCII text file (you may even edit it).
%However, you need to transfer a PDF file or any compressed or packaged
%file in binary mode when using FTP.

\label{lastpage}


\begin{thebibliography}{99}
\small \setlength{\itemindent}{-3mm} \setlength{\itemsep}{-0.5mm}
\setlength{\baselineskip}{4.5mm}

%% you can type \apj for ApJ, \aap for A&A, \apss for Ap&SS, etc. Please consult
%% the macro raa.cls. You can also find them in aasguide.tex (AASTeX for ApJ, AJ, PASP)
%% Please follow the formats of RAA's references list as demonstrated below:

\bibitem[{Antonucci~}{et~al.}(2006)]{Antonucci2006}
Antonucci, E. 2006, Space Sci. Rev., 124, 35

\bibitem[{Baker~} {et~al.}(2009)]{Baker2009}
Baker, D., Van Driel-Gesztelyi, L., Mandrini, C. H., et al. 2009,
\apj, 705, 926

\bibitem[{Banerjee~} {et al.}(2000)]{Banerjee2000}
Banerjee, D., O$^{\prime}$Shea, E., \& Doyle, J. G. 2000, Sol.
Phys., 196, 63

\bibitem[{Chen~}{et~al.}(2004)]{Chen2004}
Chen, Y., Esser, R., Strachan, L., \& Hu, Y. 2004, ApJ, 602, 415

\bibitem[{Chen~}{et~al.}(2008)]{Chen2008}
Chen, P. F., Innes, D. E., \& Solanki, S. K. 2008, A\&A, 484, 487

\bibitem[{Del Zanna~}{}(2008)]{DelZanna2008}
Del Zanna, G. 2008, A\&A, 481, L49

\bibitem[{De Moortel~}{et~al.}(2000)]{DeMoortel2000}
De Moortel, I., Ireland, J., \& Walsh, R. W. 2000, A\&A, 355, L23

\bibitem[{De Moortel~}{et~al.}(2002)]{DeMoortel2002}
De Moortel, I., Ireland, J., Hood, A. W., \& Walsh, R. W. 2002,
A\&A, 387, L13

\bibitem[{De Pontieu~}{et~al.}(2004)]{DePontieu2004}
De Pontieu, B., Erd\'{e}lyi, R., \& James, S. P. 2004, Nature, 430,
536

\bibitem[{De Pontieu~}{et~al.}(2005)]{DePontieu2005}
De Pontieu, B., Erd\'{e}lyi, R., \& De Moortel, I. 2005, ApJ, 624,
L61

\bibitem[{Golub~}{et~al.}(2007)]{Golub2007}
Golub, L., et al. 2007, Sol. Phys., 243, 63

\bibitem[{Harra~}{et~al.}(2008)]{Harra2008}
Harra, L. K., et al. 2008, ApJ, 676, L147

\bibitem[{Hassler~}{et~al.}(1999)]{Hassler1999}
Hassler, D. M., Dammasch, I. E., Lemaire, P., Brekke, P., Curdt, W.,
Mason, H. E., Vial, J.-C., \& Wilhelm, K. 1999, Science, 283, 810

\bibitem[{He~}{et~al.}(2007)]{He2007}
He, J.-S., Tu, C.-Y., \& Marsch, E. 2007, A\&A, 468, 307

\bibitem[{He~}{et~al.}(2009)]{He2009}
He, J.-S., Tu, C.-Y., Tian, H., \& Marsch, E. 2009, Adv. Space Res.,
45, 303

\bibitem[{He~}{et~al.}(2010)]{He2010}
He, J.-S., Marsch, E., Tu, C.-Y., Guo, L.-J., \& Tian, H. 2010,
A\&A, in press

\bibitem[{Kohl~}{et~al.}(2006)]{Kohl2006}
Kohl, J. L., Noci, G., Cranmer, S. R., \& Raymond, J. C. 2006,
Astron. Astrophys. Rev., 13, 31

\bibitem[{Kojima~}{et~al.}(1999)]{Kojima1999}
Kojima, M., et al. 1999, J. Geophys. Res., 104, 16993

\bibitem[{Krieger~}{et~al.}(1973)]{Krieger1973}
Krieger, A. S., Timothy, A. F., \& Roelof, E. C. 1973, Sol. Phys.,
29, 505

\bibitem[{Marsch~}{et~al.}(2008)]{Marsch2008}
Marsch, E., Tian, H., Sun, J., Curdt, W., \& Wiegelmann, T. 2008,
ApJ, 684, 1262

\bibitem[{McIntosh \& De Pontieu}{}(2009)]{McIntosh2009}
McIntosh, S. W., \& De Pontieu, B., 2009, ApJ, 706, L80

\bibitem[{McIntosh~}{et~al.}(2010)]{McIntosh2010}
McIntosh, S. W., Innes, D. E., De Pontieu, B., \& Leamon, R. J.
2010, A\&A, 510, L2

\bibitem[{Robbrecht~}{et~al.}(2001)]{Robbrecht2001}
Robbrecht,E., et al. 2001, \aap, 144, 469

\bibitem[{Sakao~}{et~al.}(2007)]{Sakao2007}
Sakao, T., et al. 2007, Science, 318, 1585

\bibitem[{Schrijver~}{}(2001)]{Schrijver2001}
Schrijver, C. J. 2001, ApJ, 547, 475

\bibitem[{Tian~}{et~al.}(2008)]{Tian2008a}
Tian, H., Tu, C.-Y.,Marsch, E., He, J.-S., \& Zhou, G.-Q. 2008a,
A\&A, 478, 915

\bibitem[{Tian \& Xia~}{}(2008)]{Tian2008b}
Tian, H., \& Xia, L.-D.  2008b, \aap, 488, 331

\bibitem[{Tian~}{et~al.}(2009)]{Tian2009}
Tian, H., Marsch, E., Curdt, W., \& He, J.-S. 2009, ApJ, 704, 883

\bibitem[{Tian~}{et~al.}(2010)]{Tian2010}
Tian, H., Tu, C.-Y., Marsch, E., He, J.-S., \& Kamio, S. 2010, ApJ,
709, L88

\bibitem[{Torrence \& Compo~}(1998)]{TorrenceCompo1998}
Torrence, C., \& Compo, G. P., 1998. Bull. Amer. Meteor. Soc. 79, 61

\bibitem[{Tu~}{et~al.}(2005)]{Tu2005}
Tu, C.-Y., Zhou, C., Marsch, E., Xia, L.-D., Zhao, L., Wang, J.-X.,
\& Wilhelm, K. 2005, Science, 308, 519

\bibitem[{Wang~}{et~al.}(1990)]{Wang1990}
Wang, Y.-M., Sheeley, N. R., Jr., \& Nash, A. G. 1990, Nature, 347,
439

\bibitem[{Winebarger~}{et~al.}(2001)]{Winebarger2001}
Winebarger, A. R., et al. 2001, ApJ, 361, 309


\end{thebibliography}
\end{document}